# Outage Probability Analysis of Mixed RF-FSO System Influenced by Fisher–Snedecor Fading and Gamma-Gamma Atmospheric Turbulence

Milica I. Petkovic, Predrag N. Ivanis, Goran T. Djordjevic

*Abstract* — In this paper, we investigate a dual-hop relaying system, composed of radio frequency (RF) and free-space optical (FSO) link. Decode-and-forward (DF) relay is employed to integrate the first RF link and the second line-of-sight FSO links. The RF channel is assumed to be subject to recently proposed Fisher–Snedecor fading model, which was shown to be convenient for modeling in realistic wireless communication scenarios. The FSO channel is affected by Gamma-Gamma distributed atmospheric turbulence. Expression for the outage probability is derived and utilized to present numerical results. Based on presented results, the effects of various RF and FSO channels parameters on the overall system performance are examined and discussed.

## I. Introduction

THE free space optics (FSO) is an emerging technology that can be utilised in contemporary communication systems due to many advantages as secure communications, high data rates, low-cost implementation etc. [1] – [4]. Utilisation of relays in FSO systems has been extensively studied in [5] – [9] with aim to increase area of coverage. It is very difficult to establish line-of-sight (LoS) between transmitter and receiver of FSO system. Because of that, the so-called mixed dual-hop relaying architecture was firstly suggested in [10]. This system consists of radio frequency (RF) link, relay station and FSO link. This architecture also enables multiple RF users to be multiplexed over a single FSO link. In order to convert electrical-to-optical signal in the relay station, subcarrier intensity modulation (SIM) can be performed.

Recent literature contains a number of studies related to the analysis of the mixed RF-FSO system performance. In [10]–[15], different system metrics have been observed considering the mixed RF-FSO system with fixed amplified-and-forward (AF) relay gain. On the other hand, the performance of the RF/FSO system with variable AF relay gain was observed in [15] – [20]. Furthermore, the mixed RF-FSO system employing decode-and-forward (DF) relay was analyzed in [21] – [24].

Inspired by those studies, in this paper, we observe the mixed RF-FSO system employing DF relay. Unlike previous studies, which observe different types of the fading scenarios over the first RF hop, in this paper we consider that the RF channel experiences recently proposed Fisher–Snedecor fading model [25]. Fisher–Snedecor fading model is proved to be appropriate to characterize the combined effects of multipath fading and shadowing in emerging wireless communication scenarios. As it was shown in [25], this distribution can be utilized to model device-to-device communication links at 5.8 GHz, in both indoor and outdoor environment scenarios. Furthermore, it can be reduced to some well-known distributions, such as the Nakagami-*m*, Rayleigh and one-sided Gaussian distribution [25] – [27]. The intensity fluctuations of the optical signal over the FSO hop are modeled by widely accepted Gamma-Gamma distribution [2] – [4]. We derive analytical expression for the outage probability and present numerical results.

The rest of the paper is organized as follows. Section II describes the system and channel model. The outage probability analysis is presented in section III. Numerical results with comments are depicted in section IV. Concluding remarks are listed in section V.

## II. System and Channel Models

We analyze mixed relaying system composed of the first RF hop and the second FSO link. The DF relay is used to link the RF and the LoS FSO hops. In the first part of communication, the RF modulated signal is transmitted from the source to the relay node via the RF link. The received electrical signal at the DF relay has the form

$$r = hs + n_r, \qquad (1)$$

where *h* is the fading amplitude of the RF link, and *s* represents the signal sent from the source node. The average transmitted electrical power is denoted with $P_s$, and $n_r$ denotes an additive white Gaussian noise (AWGN) with zero mean and variance $\sigma_r$. The DF relay performs re-transmission of signal on the optical wireless frequency via FSO link. Receiving telescope collects the optical signal at the destination node, and direct detection is performed. The optical-to-electrical signal conversion is performed by PIN photodetector. The received electrical signal is given by

Corresponding Milica I. Petkovic is with University of Novi Sad, Faculty of Technical Sciences, 21000 Novi Sad, Serbia (phone: +381-21-485-2518; e-mail: milica.petkovic@uns.ac.rs).

Predrag N. Ivanis is with the School of Electrical Engineering, University of Belgrade, Bul. kralja Aleksandra 73, 11120 Belgrade, Serbia (phone: +381-11-324-8464; e mail: predrag.ivanis@etf.rs).

Goran T. Djordjevic is with University of Nis, Faculty of Electronic Engineering, 18000 Nis, Serbia (phone: +381-18-529-424; e-mail: goran@elfak.ni.ac.rs).

$$d = P_t \eta I r + n_d, \quad (2)$$

where $I$ is the irradiance amplitude over FSO link, $\eta$ is the optical-to-electrical conversion coefficient, and $P_t$ represents the average transmitted optical power. The AWGN, with zero mean and variance $\sigma_d$, over an FSO link is denoted by $n_d$.

For the observed system, the equivalent end-to-end SNR is defined as

$$\gamma_{eq} = \min(\gamma_1, \gamma_2), \quad (3)$$

where $\gamma_1$ is the instantaneous SNR of the RF link defined as

$$\gamma_1 = \frac{|h|^2 P_s}{\sigma_r^2}, \quad (4)$$

while $\gamma_2$ is the instantaneous SNR of the FSO link defined as

$$\gamma_2 = \frac{P_t^2 \eta^2 I^2}{\sigma_d^2}. \quad (5)$$

### A. RF link

We adopt recently proposed Fisher–Snedecor statistical model to describe the effect of composite fading over the first hop. It is assumed that the scattered multipath fading is subject to a Nakagami-$m$ distribution, while the root-mean-square signal is described by variation induced by shadowing and experience an inverse Nakagami-$m$ random variable. The probability density function (PDF) of the instantaneous SNR of the RF link, $\gamma_1$, is determined as [25]

$$f_{\gamma_1}(\gamma) = \frac{m^m (m_s \mu_1)^{m_s} \gamma^{m-1}}{B(m, m_s)(m\gamma + m_s \mu_1)^{m+m_s}}, \quad (6)$$

where $\mu_1 = \mathrm{E}[\gamma_1]$ is the mean SNR with the mathematical expectation denoted by $\mathrm{E}[.]$, $m$ and $m_s$ are the fading severity and shadowing parameters, respectively, and $B(.,.)$ represents the Beta function [8, (8.384.1)]. The Fisher–Snedecor model can be reduced to the some special cases: for $m_s \to \infty$ to the Nakagami-$m$ fading model, for $m_s \to \infty, m = 1$ to the Rayleigh fading model, for $m_s \to \infty, m = 0.5$ to one-sided Gaussian distribution.

The cumulative distribution function (CDF) of the instantaneous SNR, $\gamma_1$, can be derived as

$$F_{\gamma_1}(\gamma) = \int_{-\infty}^{\gamma} f_{\gamma_1}(x) dx$$
$$= \frac{m^{m-1} \gamma^m {}_2F_1\left(m, m+m_s; m+1; -\frac{m\gamma}{m_s \mu_1}\right)}{B(m, m_s)(m_s \mu_1)^m}. \quad (7)$$

### B. FSO link

The intensity fluctuations of the received optical signal are modeled by Gamma-Gamma distribution. The PDF of the instantaneous SNR is given by [2] – [4]

$$f_{\gamma_2}(\gamma) = \frac{(\alpha\beta)^{\frac{\alpha+\beta}{2}}}{\Gamma(\alpha)\Gamma(\beta)\mu_2} \left(\frac{\gamma}{\mu_2}\right)^{\frac{\alpha+\beta}{4}-1} \times K_{\alpha-\beta}\left(2\sqrt{\alpha\beta\sqrt{\frac{\gamma}{\mu_2}}}\right), \quad (8)$$

where $\mu_2$ represents the average electrical SNR, $\alpha$ and $\beta$ are the atmospheric turbulence parameters related to the atmospheric conditions, $\Gamma(.)$ denotes the Gamma function [28, (8.310.1)] and $K_\nu(.)$ is the $\nu$-th order modified Bessel function of the second kind [28, (8.432)]. The average electrical SNR is defined as

$$\mu_2 = \frac{P_t^2 \eta^2 \mathrm{E}[I]^2}{\sigma_d^2} = \frac{P_t^2 \eta^2}{\sigma_d^2}, \quad (9)$$

since $I$ is normalized, i.e., $\mathrm{E}[I] = 1$. The atmospheric turbulence parameters $\alpha$ and $\beta$ are defined as

$$\alpha = \left(\exp\left[\left(0.49\sigma_R^2\right)/\left(1+1.11\sigma_R^{12/5}\right)^{7/6}\right]-1\right)^{-1}$$
$$\beta = \left(\exp\left[\left(0.51\sigma_R^2\right)/\left(1+0.69\sigma_R^{12/5}\right)^{7/6}\right]-1\right)^{-1}, \quad (10)$$

where $\sigma_R^2$ is the Rytov variance determined as

$$\sigma_R^2 = 1.23 C_n^2 k^{7/6} L^{11/6}, \quad (11)$$

with the wave-number $k = 2\pi/\lambda$, the wavelength $\lambda$, while $L$ is the propagation distance. The refractive index structure parameter is denoted with $C_n^2$, which determines the atmospheric turbulence strength and typically varies from $10^{-17}$ to $10^{-13}$ m$^{-2/3}$.

The CDF of the instantaneous SNR, $\gamma_2$, is given by

$$F_{\gamma_2}(\gamma) = \frac{1}{\Gamma(\alpha)\Gamma(\beta)} G_{1,3}^{2,1}\left(\alpha\beta\sqrt{\frac{\gamma}{\mu_2}}\bigg|\begin{matrix}1\\ \alpha, \beta, 0\end{matrix}\right). \quad (12)$$

## III. OUTAGE PROBABILITY ANALYSIS

In order to obtain analytical expression for the outage probability, we need to find the CDF of the equivalent end-to-end SNR defined in (3). For considered DF relying system, the CDF of $\gamma_{eq}$ is determined as

$$F_{\gamma_{eq}}(\gamma) = \Pr[\gamma_{eq} < \gamma]$$
$$= F_{\gamma_1}(\gamma) + F_{\gamma_2}(\gamma) - F_{\gamma_1}(\gamma) F_{\gamma_2}(\gamma) \quad (13)$$

where $F_{\gamma_1}(\gamma)$ and $F_{\gamma_2}(\gamma)$ are the CDFs of the SNRs over RF and FSO links, previously defined in (7) and (12), respectively.

The outage probability, defined as the probability that the instantaneous equivalent end-to-end SNR drops below

a predetermined outage protection threshold, $\gamma_{th}$, can be derived as

$$P_{out} = F_{\gamma_{eq}}(\gamma_{th}). \quad (14)$$

## IV. NUMERICAL RESULTS

In this section, we present numerical results obtained based on the analytical outage probability expression in (14). Turbulence parameters $\alpha$ and $\beta$ are defined by (10). The atmospheric turbulence strength is determined by the refractive index structure parameter in the following way: $C_n^2 = 6 \times 10^{-15}$ m$^{-2/3}$, $C_n^2 = 2 \times 10^{-14}$ m$^{-2/3}$ and $C_n^2 = 6 \times 10^{-15}$ m$^{-2/3}$ for weak, moderate and strong conditions, respectively. The value of wavelength is $\lambda = 1550$ nm. Numerical values of the Fisher–Snedecor statistical distribution are taken from [25], depending on different environment scenarios.

Fig. 1. presents the outage probability dependence on $\mu_1 = \mu_2$ for different values of fading severity and shadowing parameters. Based on values of $m$ and $m_s$ [25], it is assumed that the RF signal transmission is performed in indoor open office environment over head-to-head (H2H) and head-to-pocket channels (H2P). It can be concluded the LoS transmission ($m$=1.12, $m_s$=1.42 and $m$=0.98, $m_s$=2.03) in the RF hop reflects in better system performance compared to the non LoS transmission ($m$=1.09, $m_s$=2.25 and $m$=0.75, $m_s$=4.27), for both H2H and H2P.

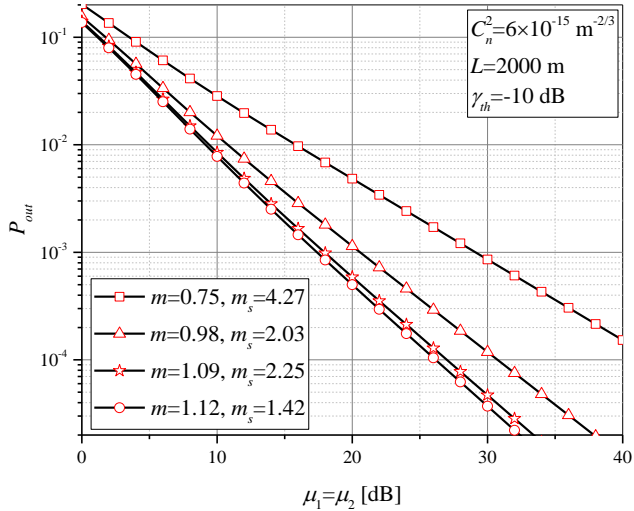

Fig. 1. Outage probability versus $\mu_1 = \mu_2$ for different values of fading severity and shadowing parameters.

Outage probability dependence on the average electrical SNR over FSO link in different atmospheric turbulence conditions is shown in Fig. 2. The RF signal transmission is accomplished in indoor office environment for H2H channel. As expected, when the refractive index structure parameter is lower, atmospheric turbulence is weaker, and FSO signal transmission is performed in better condition. Hence, overall system performance is improved when the optical wireless transmission is affected by weak atmospheric turbulence. Additionally, the RF channel conditions have lower impact on total system performance when the FSO hop is influenced by deteriorated strong turbulence. Furthermore, the increase of $\mu_2$, i.e., average transmitted optical power, leads to the existence of outage probability floor. This means that further increasing of optical power will not improve the system performance. This outage floor is different in various atmospheric turbulence conditions, but it is independent on RF channel state.

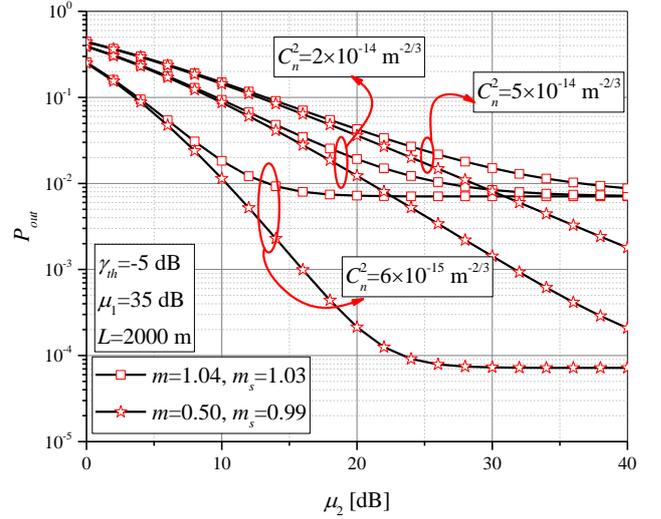

Fig. 2. Outage probability versus $\mu_2$ in various atmospheric turbulence conditions.

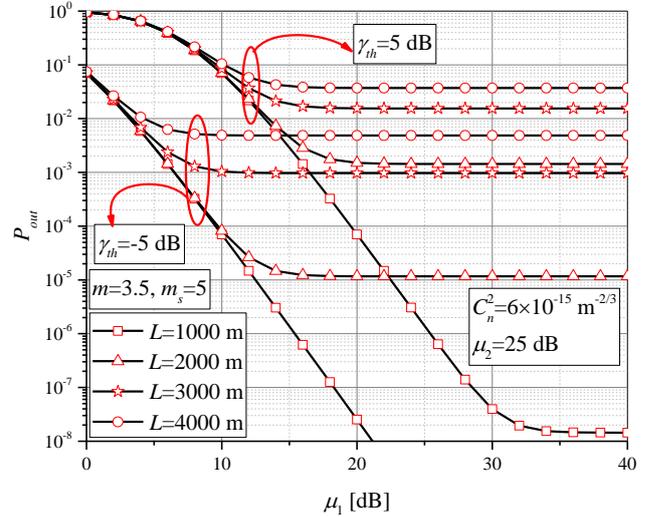

Fig. 3. Outage probability versus $\mu_1$ for different FSO link lengths.

Fig. 3. depicts the outage probability dependence on $\mu_1$ considering different FSO link length. When the optical link is longer, received optical power is reduced, and the overall system performance is worse. The existence of the outage probability floor is also noticed, meaning that the further increase in electrical signal power at the source node will not lead in improved system performance.

## V. Conclusion

The analysis of dual-hop RF-FSO relaying system employing DF relay has been presented. The first RF link is subject to newly proposed Fisher–Snedecor fading model, which was proved to be convenient for realistic wireless communication links modeling, in both indoor and outdoor scenarios. The second outdoor FSO hop is influenced by the Gamma-Gamma distributed atmospheric turbulence. The outage probability expression is derived and used to obtain numerical results. Impact of both RF and FSO channels parameters on overall RF-FSO system performance has been observed and discussed. The existence of the outage probability floor has been noticed.

## Acknowledgment

This work was supported by the Ministry of Education, Science and Technology Development of the Republic of Serbia under Grants TR-32028 and TR-32035. The work of M. I. Petkovic has received funding from the European Union Horizon 2020 research and innovation programme under the Marie Skodowska-Curie grant agreement No 734331. The work is a part of a COST action CA 16220.